\documentstyle[12pt]{article}
\newcommand{\be}{\begin{equation}}
\newcommand{\ee}{\end{equation}}
\newcommand{\bea}{\begin{eqnarray}}
\newcommand{\eea}{\end{eqnarray}}
\newcommand{\rmi}{i}
\newcommand{\rme}{\,\,{\rm e}}

\begin{document}
\begin{titlepage}
\title{\bf A Quark Transport Theory  to describe Nucleon--Nucleon Collisions
\footnote{Part of the dissertation of
          Ulf Kalmbach}
\footnote{Supported by BMFT and GSI Darmstadt}}
\author{U.\ Kalmbach, T.\ Vetter,
T.\ S.\ Bir\'{o}, and U.\ Mosel
\\Institut f\"ur Theoretische Physik
, Universit\"at Giessen\\
6300 Giessen, West Germany}
\date{}
\maketitle

\begin{abstract}
On the basis of the Friedberg--Lee model we formulate a semiclassical transport
theory to describe the
phase--space evolution of nucleon--nucleon collisions on the quark level. The
time
evolution is given by a Vlasov--equation for the quark phase--space
distribution and a
Klein--Gordon equation for the mean--field describing the nucleon as a soliton
bag. The
Vlasov equation
is solved numerically using an extended testparticle method. We test the
confinement
mechanism
and mean-field effects in $1+1$ dimensional simulations.
\end{abstract}
\end{titlepage}

\section{Introduction} \label{Intro}

In this paper we extend the established transport theories, which
describe dynamics of heavy--ion--collisions on the basis of nucleons
(\cite{CaMo90} --
\cite{Boter}), to a theory whose basic ingredients are quarks which are the
relevant degrees
of freedom at bombarding energies from some $GeV$ up to some $TeV$.

The fundamental theory for the interaction of quarks is QCD. There are
derivations of transport equations for quarks and gluons based on QCD in the
literature
\cite{Elze1}, \cite{Elze2}, \cite{Stan}. The problem is the complexity of the
resulting
equations which makes them rather tedious for practical applications. All other
approaches
start with some approximations to QCD. For bombarding energies larger than
$\approx 100
\, GeV$ QCD can be treated perturbatively. This is the motivation for many
parton cascade
models (Fritjof \cite{AnGuNi87}, Venus \cite{We89}, RQMD \cite{SoStGr89/2},
HIJING
\cite{WaGy91}, parton cascade \cite{GeMu92}).

We are interested in the low energy regime of a few $GeV$ where the QCD
coupling
constant is too large to treat QCD perturbatively. The non--perturbative QCD
effects can be
modeled by a so--called mean--field, which provides the confinement and governs
the
dynamics of the quarks.

For the description of properties of the nucleon we have many well-established
static
quark models at our disposal. As a starting point for a dynamical theory we
need a model,
which generates dynamically a surface and is able to simulate absolute
confinement.
The last point is essential because there will be excitations of quarks in a
collision which
could lead to deconfinement if the confining potential is finite. The simplest
model which
fulfills our requirements is the well--known Friedberg--Lee--soliton model with
a
field--dependent coupling constant (for a review see \cite{Wilbuch}). Starting
from this
model we will derive transport equations for quarks moving in a mean--field. In
this paper
we study the resulting model in 1+1 dimensions to check the numerical methods
and the
behavior of the model in nucleon--nucleon collisions.

Zhang and Wilets \cite{Wil} have derived transport equations based on the
Nambu--Jona--Lasinio model in order to estimate chiral symmetry effects in
heavy--ion
collisions; this  is conceptionally close to our work. However, besides the
absence of
confinement in their model, these authors do not actually perform dynamical
simulations.

This paper is organized as follows: In section \ref{FBL} we review the static
Friedberg-Lee
model and present results in one spatial dimension. In section \ref{Transport}
we derive
equations of motion for the phase--space evolution of the quarks in this model.
Section
\ref{2Sorten} discusses the extension of the usual testparticle ansatz to
include particle and
antiparticle degrees of freedom. The initialization of a stationary nucleon in
the semiclassical
approximation is discussed in section \ref{semicl}. First simulations of
nucleon--nucleon
collisions in 1+1 dimensions are presented in section \ref{collisions}.
Finally, we summarize
and conclude in section \ref{summary}.

\section{The Friedberg--Lee Model} \label{FBL}

The Friedberg--Lee soliton model in its basic version was formulated in 1977
and 1978 by
Friedberg and Lee \cite{FbLe77},\cite{FbLe78}. The quark--quark interaction in
this model
is mediated by a selfinteracting scalar $\sigma$ field. The scalar field is
interpreted as a
summation of all nonperturbative gluonic interactions between the quarks.
Because of its
color neutral nature, the $\sigma$--field can only model many--gluon--exchange.
Therefore,
the model has been extended to include
absolute color confinement by introducing a color--dielectric function
\cite{BiBiWi87}.
In this paper we use the following simplified version of this model
\cite{FaPeWi88}
\be
{\cal L} = \rmi \bar{\Psi} \left(\gamma_\mu \partial^\mu - m_0 \right) \Psi -
\bar{\Psi}
g_{eff}(\sigma) \Psi + \frac{1}{2} \left(\partial_\mu \sigma \right)^2 - U
(\sigma) ,
\label{FBLmod}
\ee
which describes quarks with a rest mass $m_0$ coupled to a scalar field
$\sigma$. This
coupling together with the mass term leads to an explicit breaking of chiral
symmetry. For
the effective coupling to the scalar field we use the form given by Fai et
al.\cite{FaPeWi88}
 (a similar coupling is discussed in
\cite{BaFoWe86})
\be
g_{eff}(\sigma) = g_0 \, \sigma_{vac}\left[\frac{1}{\kappa(\sigma)} - 1\right]
\label{ggeff}
\ee
and
\be
\kappa(\sigma) = 1 + \theta(x) x^n [nx -(n + 1)]  \label{theta}
\ee
\be
\mbox{with}\quad\quad x = \frac{\sigma}{\sigma_{vac}}\quad\quad \mbox{and}
\quad
n = 3.
\ee
It is shown in \cite{FaPeWi88} that this form of the effective coupling gives a
good
approximation to the effects of the gluon field in the chirally invariant color
dielectric
extension of the soliton--bag model. It guarantees absolute confinement,
because the quarks
acquire an infinite  effective mass $m_0 + g_{eff}(\sigma)$ in the vacuum. The
$\theta$--
function in (\ref{theta}) guarantees that the effective quark mass is larger
than the quark
rest mass and always positive ($g > 0$), which is essential for our
semiclassical treatment
of this model. Therefore we start with  the Lagrangian (\ref{FBLmod}) to
formulate a
transport theory.

For the scalar field $\sigma$, a nonlinear self-interaction $U(\sigma)$ is
assumed, which
is necessary to allow for solitonic solutions of the field. This nonlinear
potential
is parametrized as
\be
U(\sigma) = \frac{a}{2!} \sigma^2 + \frac{b}{3!} \sigma^3 +\frac{c}{4!}
\sigma^4 + B
\quad .
\ee
The free parameters of the model $g_0$, $a$, $b$, $c$, $B$ can be adjusted to
reproduce the basic properties of the nucleon, namely mass, RMS radius,
magnetic moment
and the ratio $g_V/g_A$.

Solutions of the Friedberg--Lee model have been extensively studied during the
last decade
\cite{Wilbuch},\cite{WiBiLuHe85} -- \cite{BiBiWi87}.

In this paper we restrict our studies to $1+1$ dimensions. In order to
ascertain that the
essential properties of the model are still present in $1+1$ dimensions we have
compared
the solutions of the Friedberg--Lee model in 1 spatial dimension with the
3--dimensional
results from the literature. It has turned out, that the results are
essentially identical. In
order to show this, we present a 1--dimensional result for the model
(\ref{FBLmod}) with
the effective coupling given by (\ref{ggeff}). In one space--dimension the
ansatz for the
spinless quark spinors is
\be
\Psi =\left(\begin{array}{c} u \\ \rmi	v \end{array} \right),
\ee
which gives the following equations of motion in the usual mean--field
approximation
\bea
\frac{d u}{d x} &=& -(\varepsilon + g \sigma + m_0) \, v , \\[4mm]
\frac{d v}{d x} &=&  (\varepsilon - g \sigma - m_0) \, u  , \\[4mm]
\frac{d^2}{d x^2} \sigma - \frac{d U(\sigma)}{d \sigma} &=& N \, \frac{d
g(\sigma)}{d
\sigma} (u^2 - v^2) \quad ,
\eea
with the occupation number $N=3$ for the valence orbital. A typical solution
for the
parameters
$a=222.91$, $b=-5347.05$, $c=38610$, $g_0=1$ with a total energy of $931 MeV$
and an RMS radius of $0.56 fm$ is shown in figs. \ref{Abb1} and \ref{Abb2}.

\section{Derivation of the Transport Equation} \label{Transport}

The aim of this section is to derive equations of motion which describe the
phase--space
evolution of a moving nucleon. We therefore follow the spirit of the well known
RBUU model \cite{CaMo90,Blattel}, which has been applied very successfully to
heavy ion
collisions \cite{Blattel}. The following steps are similar to the derivation in
\cite{EVG,Blattel}. Starting with a static model for the nucleon, the
Friedberg--Lee model,
we derive an equation for the time evolution of the Wigner--function, which is
the quantum
mechanical analog of the classical phase--space density. The Wigner--function
is defined as
\be
W^{r_i , r_j} (x,p) = \frac{1}{(2 \pi )^4} \int d^4 R \rme^{- \rmi p_{\mu}
R^{\mu}}
                                 \bar{\Psi}^{r_i} (x + \frac{R}{2})  \otimes
                                 \Psi^{r_j} (x - \frac{R}{2}) \quad ,
\ee
where the indices $r_i , r_j $ label internal degrees of freedom, such as
color, flavor, etc; in
the following they are suppressed. Densities and particle number can be easily
calculated by
integrating over the Wigner--function :
\bea
\rho (x) &=& \int d^4 p \, \, tr \left( \gamma_0 W(x,p) \right) \\
\rho (p) &=& \int d^4 x \, \, tr \left( \gamma_0 W(x,p) \right)  \\
N &=& \int d^4 x \int d^4 p \, \, tr \left( \gamma_0 W(x,p) \right) \quad .
\eea
In general, expectation--values of one--particle operators $\hat{O}$ are given
by
\be
\left< \hat{O} \right> = \int d^4 x \int d^4 p \, \, tr ( \hat{O} W(x,p) )
\quad .
\ee
The equation of motion for the Wigner--function can be derived by calculating
\be
\left[ \gamma_{\mu} \left( \partial^{\mu}_{x} - 2 \rmi p^{\mu} \right) \right]
W(x,p) =
\frac{2}{(2 \pi)^4} \int d^4 R \rme^{- \rmi p_{\mu} R^{\mu}}
\bar{\Psi}(x_1)  \otimes  \gamma_{\nu} \partial^{\nu}_{x_2} \Psi(x_2) \quad ,
\label{Wig1}
\ee
where
\be
x_1 = x + \frac{R}{2} \quad , \quad x_2 = x - \frac{R}{2} .
\ee
On the r.h.s. of equation (\ref{Wig1}) one can use the Dirac--equation of the
Friedberg--Lee--model to replace $ \gamma_{\nu} \partial^{\nu}_{x_2} \Psi(x_2)
$ which
gives
after some algebraic transformations:
\bea
&&\left[ \gamma_{\mu} \left( \hbar \partial^{\mu}_{x} - 2 \rmi p^{\mu} \right)
\right]
W(x,p) =
\nonumber \\
&&- 2 \rmi \rme^{\frac{\rmi}{2} \hbar \partial^p_{\mu} \partial^{\mu}_{x}}
\left[  m_0 + g \sigma (x) \right]  W(x,p) \quad . \label{Wig2}
\eea
The derivative $\partial_\mu^p$ acts on the Wigner--function and
$\partial_x^\mu$ acts on
the
field $\sigma(x)$. The factors $\hbar$ are inserted explicitly. Up to now,
equation (\ref{Wig2})
together with the equation for the mean field
\be
\partial_\mu \partial^\mu \sigma - \frac{d U(\sigma)}{d \sigma} =
\frac{d g(\sigma)}{d \sigma} \int d^4 p \, \, tr  W(x,p)
\ee
is equivalent to the equations of motion for the Friedberg--Lee model in the
mean--field approximation. In the semiclassical approximation, one expands the
Wigner--function and the exponential function in equation (\ref{Wig2}) in
orders of
$\hbar$:
\bea
W &=& W_0 + \rmi \hbar W_1 + \cdots \\ \nonumber
 \rme^{\frac{\rmi}{2} \hbar \partial^p_{\mu} \partial^{\mu}_{x}} &=& 1 +
\frac{\rmi}{2}
\hbar \partial^p_{\mu} \partial^{\mu}_{x} + \cdots \quad .
\eea
This expansion is expected to converge the better the less the fields vary over
the Compton
wavelength of the quarks.
In lowest order $\hbar$ this leads to the equation
\be
\left( \gamma_{\mu} p^{\mu} - m^{\ast} \right) W_0(x,p) = 0 \quad ,
\label{effective}
\ee
with $ \quad m^{\ast}= m_0 + g_{eff} (\sigma (x)) \quad $.

This is the well known mass--shell constraint.
In first order $\hbar$ one has:
\be
\left[ \gamma_{\mu} \partial_x^{\mu} + \partial_{\mu}^p \partial_x^{\mu}
m^{\ast}(x)
\right]
W_0(x,p) = 2 \left[ \gamma_{\mu} p^{\mu} - m^{\ast} \right] W_1(x,p)
\label{Wig3}
\quad .
\ee
We now take the trace on both sides of this equation. If one requires baryon
current
conservation in lowest order in $\hbar$, $W_1$ has to fulfill a
constraint so that the trace of the r.h.s. of equation (\ref{Wig3}) vanishes.
We
are then left with an equation of motion for $W_0$, the well known
Vlasov--equation.

All equations are so far derived for $3+1$ dimensions. From now on we study the
model in
$1+1$ dimensions in order to test its main features and the numerical methods
for its
solution. The generalization to $3+1$ dimensions will be discussed in a future
work. In
$1+1$ dimensions we introduce Dirac-Matrices $\Gamma_\mu$, which have to
fulfill the
usual anti--commutation relations:
\be
\left\{ \Gamma_{\mu} , \Gamma_{\nu} \right\} = g_{\mu \nu} \quad \mu , \nu \in
\left\{ 0 ,
1
\right\}
\ee
\be
\mbox{with} \quad g_{\mu \nu} = \left( \begin{array}{cc} 1 & 0 \\ 0 & -1
\end{array}
\right).
\ee
One possible representation for the $\Gamma$--matrices is
\be
\Gamma_0 = \left( \begin{array}{cc} 1 & 0 \\ 0 & -1 \end{array} \right)  ,
\quad
\Gamma_1 = \left( \begin{array}{cc} 0 & 1 \\ -1 & 0 \end{array} \right)  .
\ee
One also defines the product
\be
\Gamma_2 = \Gamma_0 \Gamma_1 =  \left( \begin{array}{cc} 0 & 1 \\ 1 & 0
\end{array}
\right) ,
\ee
which is analogous to $\gamma_5$ in 3+1 dimensions. \\
The relations
\be
\left\{ \Gamma_2 , \Gamma_{\mu} \right\} = 0 \quad , \quad \Gamma_0^2 = I
\quad ,
\quad \Gamma_1^2 = - I \quad , \quad \left[ I ,\mbox{and} \Gamma_{\mu} \right]
= 0 ,
\ee
hold for these matrices, where $I$ is the $2 \times 2$ unit matrix.

In the 1--dimensional notation the Vlasov--equation can be written in the
following form:
\be
tr \, \left( \left[ \Gamma_0 \partial_t + \Gamma_1 \partial_{x'} + I
(\partial_t m^{\ast}
\partial_{E} -
\partial_{x'} m^{\ast} \partial_{p^1} )  \right]
W_0(x,p) \right) = 0  \label{Wig4} ,
\ee
with $x=(t,x^1)$ and $p=(E,p^1)$.

This equation incorporates energy-- and particle--number conservation; it
approximates
the quantum--mechanical solution only if the fields vary slowly, a condition
which we will discuss later.

In the following we deal with $W_0$ only and suppress the subscript. In view of
the
mass--shell constraint eq. (\ref{effective}), one takes the following ansatz
for the Wigner matrix:
\be
W(x,p) = \left( \Gamma_\mu p^\mu + m^{\ast} \right) f(x,p)  \quad ,
\ee
where $f(x,p)$ is a scalar function of $x=(t,x^1)$ and $p=(E,p^1)$.
This is equivalent to the usual spinor decomposition neglecting the
pseudoscalar part.
Insertion of this ansatz in equation (\ref{Wig4}) finally gives the
Vlasov--equation:
\be
\left[ p_{\mu} \partial_x^{\mu} + m^{\ast} \partial_x^{\mu} m^{\ast}
\partial_{\mu}^p
\right]
f(x,p) = 0  \label{Vlasov} .
\ee

The mass--shell constraint
\bea
\left( \Gamma_\mu p^\mu - m^{\ast} \right) W(x,p) &=& \left( \Gamma_\mu p^\mu -
m^{\ast}
\right)  \left( \Gamma_\mu p^\mu + m^{\ast} \right) f(x,p) =
\nonumber \\
&=&  \left( p^2 - {m^{\ast}}^2 \right)
f(x,p) = 0
\eea
will be fulfilled by a using a Dirac $\delta$--function $\delta \left(
p^2-{m^\ast}^2 \right)$ in
$f(x,p)$.
For the numerical treatment of this equation we use the so called testparticle
ansatz for the
scalar phase--space distribution function $f(x,p)$ \cite{Wang}:
\be
f(x,p) = \delta \left( p^2 - {m^{\ast}}^2 \right) \theta(E) \sum_n \delta (x -
x_n(t) ) \delta(
p^1 -
p^1_n(t) ) \quad . \label{Testpart1}
\ee
The $\theta$--function expresses the restriction to positive energy states.
This
approximation will be discussed in detail in section \ref{2Sorten}. Inserting
this ansatz into
the Vlasov--equation (\ref{Vlasov}) shows that the so called testparticles have
to move like
classical particles according to Hamilton equations of motion:
\bea
\dot{x_n} &=& \frac{p^1_n}{E_n} \label{Test1} \\
\dot{p}^1_n &=& - \frac{m^{\ast}_n}{E_n} \partial_x m^{\ast}_n  \quad .
\label{Test2}
\eea
with
\be
E_n = \sqrt{{p^1_n}^2 + {m^\ast}^2(x_n)}
\ee

\section{Extended testparticle ansatz} \label{2Sorten}

The usual testparticle ansatz (\ref{Testpart1}) which is restricted to positive
energy
solutions implies:
\bea
\int d E \, tr W(x,p) &=& W_{scalar}(x,p^1) = \frac{m^\ast}{E(x,p^1)}  f(x,p^1)
\label{scalar1} \\
\int d E \, tr \left( \Gamma_0 W(x,p) \right)  &=& W_{baryon}(x,p^1) =   f
(x,p^1)
\label{baryon1}
\quad.
\eea
Therefore, at an initial time instant a given baryon--density can be reproduced
by using
eq. (\ref{baryon1}). The scalar density is then given by eq. (\ref{scalar1}).
In nuclear transport theories
the rest mass of the testparticles is of the order of $1 GeV$. In this case
this ansatz
works quite well. On the quark level the rest mass of the quarks is of the
order of $5 - 10
MeV$, which is reflected in the fact that the lower component of the quantum
mechanical
wave functions is nearly as large as the upper component (see fig.\
\ref{Abb2}).  A
projection on negative energy states is not negligible. It is therefore
necessary to extend the
testparticle ansatz in order to fit scalar and baryon density simultaneously.
The importance
of the scalar density is obviously clear because it determines the dynamics of
the system
through its coupling to the $\sigma$--field,  whereas the baryon density
determines the charge
radius of the nucleon.

Our suggested extension of (\ref{Testpart1}) is given by
\be
f(x,p) = f_{pos}^{particle}(x,p) \theta(E) +
\left( 1 - f_{neg}^{hole}(x,p) \right) \theta(-E),
\ee
where
$f_{pos}^{particle}$ represents the occupation of positive energy particle
states and \\
$(1-f_{neg}^{hole})$ the occupation of negative energy hole states relative to
the filled
Dirac--sea. With this ansatz the Wigner--function is written as

\bea
W (x,p) &=& \left( \Gamma_{\mu} p^{\mu} + m^{\ast} \right)
\frac{2 \pi \delta(E - \omega)}{2 |E|} f_{pos}^{particle}(t,x,p) + \nonumber
\\[4mm]
&& + \left( \Gamma_{\mu} p^{\mu} + m^{\ast} \right) \theta(E)
\frac{2 \pi \delta(E + \omega)}{2 |E|} \left( 1 - f_{neg}^{hole}(t,x,p) \right)
\theta(-E) ,
\nonumber
\label{2Sorten1}
\eea
$\mbox{with} \quad \omega = \sqrt{p^2 + {m^{\ast}}^2}$, and
$x$ and $p$ denote the second component of the 2 vectors
$x^\mu$,$p^\mu$.
The energy dependence of the on-shell distribution functions $f(t,x,p)$ is
implicit.

Performing the energy integration we end up with:
\bea
\int \frac{d E}{2 \pi} W(t,x,E,p) &=&
\frac{ \Gamma_0 \omega - \Gamma_1 p + m^{\ast}}{2 \omega}
f_{pos}^{particle}(t,x,p)
+ \nonumber
\\[4mm]
&& + \frac{ - \Gamma_0 \omega - \Gamma_1 p + m^{\ast}}{2 \omega}  \left( 1 -
f_{neg}^{hole}(t,x,p)
\right) \quad ,
\eea
where $\omega$ is the on--shell energy. Defining
\bea
f_{pos}^{particle}(t,x,p) &=& f(t,x,p) \\[4mm]
\left( 1 - f_{neg}^{hole}(t,x,p) \right) &=&  \bar{f}(t,x,-p)
\eea
this can be rewritten as
\be
\int \frac{d E}{2 \pi} W (t,x,p,E)= \frac{ \Gamma_0 \omega - \Gamma_1 p +
m^{\ast}}{2 \omega}  f(t,x,p) \nonumber -
\frac{	\Gamma_0 \omega + \Gamma_1 p - m^{\ast}}{2 \omega}  \bar{f}(t,x,-p) .
\ee

The meaning of	$f$ and $\bar{f}$ becomes more clear if one calculates the
scalar and the baryon
density:
\bea
\rho_s (x) &=& \int dE dp \, \, tr W = \int dp	\frac{M}{E} \left( f + \bar{f}
\right) \nonumber \\
\rho_B (x) &=& \int dE dp \, \, tr \left( \Gamma_0 W \right) = \int dp \left( f
- \bar{f}
\right) .
\eea
While $f$ und $\bar{f}$ add up in calculating the scalar density, their
difference yields the baryon density.
In view of this $f$ can be interpreted as particle distibution function whereas
$\bar{f}$ is the
anti particle distribution function.

For a given Wigner--function $W$  $f$ and $\bar{f}$ will then be constructed
according to
\bea
f(x,p) &=& \frac{1}{2} \left( \frac{\omega}{m^{\ast}} W_{scalar} + W_{baryon}
\right)
\nonumber
\\
\bar{f}(x,p) &=& \frac{1}{2} \left( \frac{\omega}{m^{\ast}} W_{scalar} -
W_{baryon}
\right) \quad ,
\label{construct}
\eea
with
\be
W_{scalar} = \int dE \, \, tr W(t,x,E,p)  \quad , \quad
W_{baryon} = \int dE \, \, tr \left( \Gamma_0 W(t,x,E,p) \right)
\ee
and $ \quad \omega = \sqrt{p^2 + {m^{\ast}}^2}$ .

For the numerical realization we initialize $N_1$ testparticles according to
the distribution
$f(x,p)$ and $N_2$ testparticles according to $\bar{f}(x,p)$, where
\be
N_1 = \frac{I_1}{I_1 + I_2} \quad , \quad  N_2 = \frac{I_2}{I_1 + I_2}
\ee
\bea
\mbox{with} \quad I_1 &=& \int dx dp \, \,f(x,p) \nonumber \\
\mbox{and} \quad I_2 &=& \int dx dp \, \, \bar{f}(x,p) .
\eea
All testparticles follow the Hamiltonian equations of motion (\ref{Test1}),
(\ref{Test2}). This can be
easily understood, because in a scalar field particles and antiparticles feel
the same forces.
The baryon density and the scalar density are given by
\bea
\rho_B (x) &=& \frac{1}{\tilde{N}} \sum_{n=1}^{N} \delta (x - x_n) \, b_n \\
\rho_s (x) &=& \frac{1}{\tilde{N}} \sum_{n=1}^{N} \frac{m_n^{\ast}}{E_n}
\delta (x - x_n),
\eea
where $N=N_1 + N_2$,  $\tilde{N}=N_1-N_2$,
\bea
\mbox{and} \quad && b_n = \left\{ \begin{array}{rc} 1 & \mbox{if testparticle
from} \, \,f
\\

-1 & \mbox{if testparticle from} \, \,
\bar{f}.
						     \end{array} \right. \nonumber
\eea
On a spatial grid with spacing $\Delta x$ the $\delta$--functions are evaluated
as

\be
\delta (x - x_n ) = \left\{ \begin{array}{cl} 1/{\Delta x}	&\quad \mbox{if}
\quad x_n \in \left[ x-
\frac{\Delta x}{2} ,
 x + \frac{\Delta x}{2} \right]  \\

0 &\quad \mbox{otherwise}.  \end{array} \right.	\label{Deltaf}
\ee
With this extended testparticle ansatz, it is possible to reproduce a given
baryon density and
a scalar density independent of each other. In the se\-mi\-clas\-si\-cal
approximation the
phase--space evolution of the Friedberg--Lee model is determined by the
equations of motion
for the testparticles (\ref{Test1}) and (\ref{Test2}) and by the mean--field
equation by
\be
\partial_\mu \partial^\mu \sigma - \frac{d U(\sigma)}{d \sigma} =
\frac{d g(\sigma)}{d \sigma} \rho_s(x)
\label{Feld1} .
\ee

\section{A semiclassical nucleon} \label{semicl}

We first tried to initialize testparticles as described in the last section to
reproduce the
baryon and scalar phase--space distributions calculated from the quantum
mechanical
solution (fig. \ref{Abb2}).  Performing the time evolution of the testparticles
and the
$\sigma$--field, it turned out that the nucleon initialized in this way is not
a stable solution
of the semiclassical equation of motion (\ref{Wig4}).
This failure indicates that the conditions of weakly varying fields necessary
for the
derivation of the transport equation are actually not met. Similar difficulties
show up when
we calculate the semiclassical energy--distribution from the quantummechanical
solution.
The latter, is of course, a $\delta$--function whereas the former turns out to
be a broad
distribution.

We therefore drop now the connection between the semiclassical and the
quantummechanical solution and from now on work consistently in the
transporttheoretical
framework. In doing so we take the potential function $U(\sigma)$ as an
effective
classical potential that incorporates all higher order corrections to the
quantummechanical one and determine it by adjusting the properties of the
semiclassical solution to empirical properties.

First, we look for solutions of the stationary Vlasov--equation.
A stationary solution can be characterized by
\be
\partial_t f(x,p) = 0 \quad \mbox{and} \quad \partial_t m^{\ast} = \partial_t
\,
g_{eff}(\sigma (t)) =
0 \quad .
\ee
The Vlasov--equation reduces to
\be
\frac{E}{m^{\ast} \partial_x m^{\ast}} \partial_x f(x,p) = \frac{E}{p}
\partial_p f(x,p)
\quad ,
\ee
where $E = \sqrt{p^2 + {m^{\ast}}^2}$. Every function $f(E)$ fulfills this
equation. This
means that every distribution function which does not explicitly depend on
position and
momentum, but only on the combined quantity energy is a stationary solution of
the Vlasov-
-equation. For the initialization of a stable nucleon we have to find a
distribution which
reproduces the properties of the nucleon.

We start with a functional ansatz:
\be
f_{total}(x,p) = f(E) + \bar{f}(E)
\ee
where
\bea
f(E) &=& f_0 \rme^{- k E^2} \nonumber \\
\bar{f}(E) &=& \bar{f}_0 \rme^{- \bar{k} E^2}
\eea
are two Gaussians in energy with four free parameters; the amplitudes and the
widths.
The soliton solution is then calculated by starting with a Wood-Saxon shape for
the
$\sigma$--field. With the Gaussian ansatz we can also calculate energy and
density contributions:
\bea
\rho_s &=& \int dp \frac{M}{E} \left(   f_0 \rme^{- k E^2} +  \bar{f}_0 \rme^{-
\bar{k}
E^2}
\right),  \nonumber \\
\rho_B &=& \int dp  \left(   f_0 \rme^{- k E^2} -  \bar{f}_0 \rme^{- \bar{k}
E^2}
\right),  \nonumber \\
E_{quark} &=& \int dp \int dx E  \left(   f_0 \rme^{- k E^2} +  \bar{f}_0
\rme^{- \bar{k}
E^2}
\right) ,\nonumber \\
E_{\sigma} &=& \int dx \left( U(\sigma) + \frac{1}{2} \left( \frac{d
\sigma}{dr} \right)^2
\right) , \nonumber  \\
E_{total} &=& 3 E_{quark} + E_{\sigma},  \label{Eklass}
\eea
\[
\mbox{with} \quad E = \sqrt{p^2 + {m^{\ast}}^2 } \quad .
\]
The $\sigma$--field is then calculated selfconsistently with the scalar
density. The
parameters are varied to give reasonable values for the energy and RMS-radius
of the
nucleon. In principle also the parameters for the potential could be varied  to
find the
optimal parameter set for this 1+1 dimensional case, but this was not done
here.

In fig. \ref{SV1} we show a typical solution.
Baryon density and the  $\sigma$--field are similar to the quantum mechanical
solution. The
baryon density is volume centered. The shape of the scalar density is quite
different from
that of the baryon density and surface--peaked, reminiscent of the SLAC bag
solutions to the
Friedberg--Lee soliton bag model \cite{FbLe77}. The surface--peaking is here a
clear
consequence of the strong increase of the coupling constant $g$ towards the
surface. The
maxima at the surface are stabilizing, because the testparticle mass increases
at the surface
which results in a deceleration, meaning that the testparticles stay longer in
the surface region.
Fig. \ref{stability1} and \ref{stability2} show the stability of the
semiclassical nucleon in the
time evolution. The stability is almost perfect.

\section{The boost} \label{boost}

After the discussion of a stable initialization for a nucleon we now describe
the Lorentz--
boost of a given testparticle distribution, necessary for the preparation of
the initial state of
a collision.
The Lorentz--boost has to give the correct transformation properties for the
baryon density,
the scalar density, the $\sigma$--field and the Lorentz scalar distribution
functions $f$ and
$\bar{f}$.

We initialize the field and distribution functions at a time $t' = 0$ in the
moving frame. The
transformation is therefore given by:
\bea
&t'=0 , \qquad  \qquad &x'=\frac{x}{\gamma} \\
&p'=\gamma (p - \beta E), \qquad  \qquad &E'= \gamma (E - \beta p)
\eea
To calculate the $\sigma$-field and the distribution functions $f$ and
$\bar{f}$ at the time $t'=0$, we need to know the
configuration in the rest frame for different times $t$. In this frame,
however, the nucleon is
stationary, i.\ e.\ $f(t) = f(0)$ and $\sigma(t) = \sigma(0)$ for all times
$t$. Therefore, we
can set $t=0$.

The assumption that the distribution function is a scalar implies that
\be
f'(x',p',t'=0) = f(x,p,t=0) = f\left( \gamma x' , \gamma (p'+\beta E'),t=0
\right)
\ee
Inserting the testparticle ansatz:
\be
f(x,p,0) = \sum_n \delta (x - x_n(0) ) \delta (p - p_n(0) )
\ee
we get
\be
f'(x',p',0) = \sum_n \delta \left( \gamma x' - x_n(0) \right) \delta \left(
\gamma(p' + \beta E') -
p_n(0) \right) \quad .
\ee
We express this sum using the coordinates in the moving system:
\be
f'(x',p',0) = \sum_n (1 - \beta v_n )  \delta (x' - x_n{}'(0))  \delta (p' -
p_n{}'(0))
\label{Boost1}
\ee
\be
\mbox{with} \qquad x_n{}' = \frac{x_n}{\gamma} \qquad \mbox{and} \qquad
p_n{}' = \gamma (p_n - \beta E_n) \label{Boost2} \quad .
\ee
In addition to the usual transformation of the testparticle coordinates and
momenta often
used in the literature we obtain a factor $ (1 - \beta v_n )$ in the
testparticle sum which
results from the transformation of the arguments in the $\delta$--functions.

As a check for the expression for the boost (\ref{Boost1}), we calculate the
transformed
densities:
\bea
\rho_B{}'(x') &=& \int f'(x',p') dp' =  \sum_n \delta(x - x_n) \gamma (1 -
\beta v_n) =
\gamma
(\rho_B(x)  - \beta j_1(x)) \nonumber \\
\rho_s{}'(x') &=& \int \frac{m^{\ast}{}'}{E'} f'(x',p') dp' =  \sum_n \delta(x
- x_n)
\frac{m^{\ast}_n}{E_n} = \rho_s(x) \quad .
\eea
Here, $j_1$ denotes the baryon current, which vanishes for a stable nucleon at
rest. The densities
have the correct transformation properties. The antiparticle distribution
$\bar{f}'(x',p')$ will
be correspondingly transformed.

If one inserts the testparticle sum (\ref{Boost1}) into the Vlasov--equation,
the resulting
testparticle equations contain terms generated by the momentum derivatives
acting on the
factor $(1 - \beta v_n)$. To avoid this change, we use a new testparticle
distribution with
coordinates $\hat{x} , \hat{p}$, which have to fulfill:
\be
\sum_n \delta (x' - \hat{x}_n{}')  \delta (p' - \hat{p}_n{}')
= \sum_n (1 - \beta v_n )  \delta (x' - x_n{}')  \delta (p' - p_n{}') \quad .
\ee
This can be guaranteed, if the new testparticles located at $\hat{x} , \hat{p}$
 are
distributed according to the modified distribution function in the rest frame
\be
f(\hat{x},\hat{p}) = (1 - \beta v) f(x,p) \label{Boost3} \quad .
\ee
Finally, a moving nucleon has to be initialized in the following way:
The testparticle coordinates are distributed according to (\ref{Boost3}) and
the
transformation to
the moving coordinates is given by (\ref{Boost2}). The $\sigma$--field remains
unchanged
with
the transformed $x$ coordinates.

\section{Collisions} \label{collisions}

To simulate collisions of two nucleons, we initialize semiclassical nucleons as
described in
section \ref{semicl} and transform them to the center of mass system by
applying the boost
of section \ref{boost}. For each of the nucleons the center of the boosted
phase--space
distribution in momentum is shifted to $p_{c}{}'=\gamma \beta m_0$; due to the
small
quark rest mass, it stays close to zero momentum. This means that the boosted
phase--space
distributions always overlap in momentum space around $p \approx 0$. The
boosted phase--space
 distribution is symmetric in $x$ direction. In momentum direction one half of
the
distribution is contracted and the other is stretched, because of the
factor $(1-\beta v_n)$ in (\ref{Boost1}). This leads to a very asymmetric shape
in momentum space.

All simulations which are shown in this paper are calculated using the
testparticle method.
The testparticle propagation and the mean field time evolution are calculated
using
predictor--corrector techniques. As a check we also integrated the Vlasov
equation numerically on a
phase--space grid using an alternating direction implicit procedure (ADIP)
which leads to the
same results. The direct numerical integration can be performed only in the 1+1
dimensional
case for reasons of computing time. For the extension to 3 spatial dimensions
the
testparticle method is clearly superior for practical applications. All
calculations are done in
the c.\ m.\ system. In the simulations presented here we use 20000
testparticles for each
nucleon. This relatively large number in the 1--dimensional case is necessary
to obtain good
statistics for the calculation of the scalar density. Due to the increasing
coupling--constant
at the surface of the soliton, the $\sigma$--field is very sensitive to small
changes of the
density in this region which requires good spatial resolution that cannot be
obtained by the
rather coarse--grained smearing methods.

First, we have tested the stability of one moving nucleon. It is perfectly
stable for times
($\approx 20 fm/c$) larger than the typical collision times ($\approx 1 fm/c$).
The time evolution of the quark
distibution and the sigma field is shown in fig.\ \ref{stability1} and fig.\
\ref{stability2},
respectively.

We have calculated collisions for bombarding energies of $1 - 25 \, GeV$ in the
laboratory frame.
In the simulations we initialize two nucleons which are well separated in
space; the
propagation of a few $fm$ until the collision takes place ensures that we have
stable
incoming nucleons. For a boost velocity of $\beta > 0.9$ the collision time
is of the order of $1 \, fm/c$. To be sure to observe stable outgoing nucleons,
we
continue the calculation for $5 - 10 \, fm/c$ after the collision.

For very low bombarding energies ($T_{lab} \approx 1 GeV$) we observe a fusion
of the
two incoming bags. This is due to the slow motion of the nucleons which leads
to a long
overlap time. The testparticles are decelerated and loose kinetic energy which
results in a
gain of potential mean--field energy. The result is one extended bag with
constant density
inside. Due to the mixing of testparticles this bag does not break up into
separate bags. The
fusioned bag reaches a maximal extension and shrinks again converting potential
energy
back to kinetic testparticle energy. This behaviour is shown in fig.\
\ref{Fige1}. The
oscillation is undamped because there is no mechanism for energy loss in the
model. We
note that Schuh et. al. \cite{Schuh} estimate the potential for
nucleon--nucleon scattering in
the Friedberg--Lee soliton bag model and also end up with an attractive
potential which may be
responsible for the fusion that we observe for slowly moving bags. However, the
surface
energy in our 1+1 dimensional case is different from the real 3+1 dimensional
world so that
it is not sure that the fusion effect will survive in 3 spatial dimensions.

In collisions with bombarding energies $E_{lab}$ from $5 - 25 \, GeV$ the
nucleons are
transparent on the mean--field level. The phase--space evolution of the baryon
distribution
($f - \bar{f}$) is shown in fig.\ \ref{coll1} for a bombarding energy $E_{lab}
= 15 \, GeV$.
Since the scalar field acts in the same way on testparticles and
antitestparticles, the
distribution functions $f$ and $\bar{f}$ show the same behaviour.

 In fig.\ \ref{coll2} we display the corresponding $\sigma$--field and the
scalar source
density. In the moment the two nucleons collide, the scalar density pushes the
$\sigma$--
field barrier down. The $\sigma$--field overshoots to negative values (timestep
$t=1.1 \,
fm/c$), but due to the $\theta$--function in the effective quark--$\sigma$
coupling this does
not lead to negative effective masses. The field decouples from the density in
the region
where it is negative and oscillates up again (timestep $t=1.2 \, fm/c$).
In the following time--step ($t=1.3 \, fm/c$) the $\sigma$--field comes
actually very close to
the vacuum value. Here the coupling to the quarks then becomes large as shown
by the
large values of the scalar density. The quarks thus become very massive and
stay confined.
Remembering the physical motivation for the ingredients of our model we may say
that at
this point the explicit gluon--exchange forces take over from the color
background field.
The correlation then swings back to the background field thus reestablishing
the soliton.
The velocity of the nucleons is too large to allow for any significant
one--body dissipation
during the short overlap time. The result is a dip in the density in the moment
the nucleons
have passed through each other (timestep $t=1.6 \, fm/c$). The $\sigma$--field
then comes
close to the vacuum value and the two solitons separate (timestep $t=1.9 \,
fm/c$). If we
continue the simulation, the small deformations, which can be seen in
phase--space at  $t=1.9
\, fm/c$, vanish and we end up with two stable nucleons. The time evolution of
the energy is
shown in fig.\ \ref{Fige2}. There is no significant exchange of energy from the
testparticles
to the mean--field. This is most probably due to the restricted geometry in the
present
version of our model. A realistic surface in a $3+1$ dimensional calculation
would clearly
lead to a larger coupling between mean--field modes and the quark motion.

We conclude from these simulations that the model is able to describe moving
quark
systems and absolute confinement in a collision of nucleons. Our model is in
this sense a
basis for extensions in the direction of direct quark--quark collisions, which
we will put on
top of the stable colliding nucleons we describe so far.

\section{Summary and Conclusion} \label{summary}

In this paper we have formulated a transport theory on the quark level which
includes
nonperturbative aspects of QCD. This is done by including a soliton mean--field
which
governs the dynamics. As a starting point we chose a phenomenological quark
model for
the nucleon, namely the Friedberg--Lee model. We have derived equations of
motion for the
time evolution of the Wigner--function which leads in the semiclassical
expansion to the
well known Vlasov--equation combined with a mass--shell constraint. For the
Wigner--
function we have used a testparticle ansatz which results in classical
equations of motion for
the testparticles.

Considering different extensions to the original Friedberg--Lee model it has
turned out that
the formulation with an effective quark--$\sigma$ coupling, which comprises
non--color--singlet
many--gluon--exchange effects, can be used to model absolute confinement on the
testparticle
level. Inside the dynamically generated bag the quarks are nearly massless
(`asymptotic
freedom').

The usual testparticle ansatz was extended including also negative energy
states which was
necessary because of the small rest mass of the quarks. With the extended
ansatz it is
possible to describe scalar and baryon density independently of each other.

For the initialization of a stable nucleon we have used a consistent solution
of the stationary
Vlasov--equation. It was shown that any distribution function which depends
only on
energy and not explicitly on position and momentum is a solution of the
stationary Vlasov--
equation. After the discussion of the semiclassical nucleon we have described
how to boost
the initialized nucleon  to give boosted distribution functions with the
correct Lorentz
transformation properties.

Having all these ingredients, we have performed some first collisions in 1+1
dimensions in
order to test the numerical methods and the behavior of the model. It has
turned out that on
the pure mean field level the nucleons are totally transparent for high
bombarding energies.
The confinement is realized for each nucleon separately. For very low
bombarding energies
($\le 1 GeV$ in the lab.\ frame) we observe a fusion of the bags with a
following oscillation
of the six	quark bag. This may be an artifact of the
1--dimensional treatment.

At the moment we are extending our model to 3+1 dimensions. This is essential
to allow for
non--central collisions and sidewards flow. The next step is then to include a
direct quark--
quark collision term and cross sections for particle production.

\newpage

{\large List of figures}
\begin{figure}[h]
\caption{\label{Abb1}
Baryon density (dashed), scalar density (dotted) and $\sigma$--field (solid)
for a typical
solution
of the Friedberg--Lee model with effective coupling in 1 dimension. (The
$\sigma$--field is
in
arbitrary units.) }
\end{figure}
\begin{figure}[h]
\caption{\label{Abb2}
Upper component $u$ (solid) and lower component $v$ (dashed) of the quark
spinor for
the
1--dimensional solution of the Friedberg--Lee model with effective coupling }
\end{figure}
\begin{figure}[h]
\caption{\label{SV1}
Typical solution of the stationary Vlasov--equation with the ansatz described
in the text:
baryon
density (dashed), scalar density (dotted) and $\sigma$--field (solid). The
parameters are
$a=245.75$, $b=-5614.398$, $c=38610$, $f_0=7.527$, $k=2.0$, $\bar{f}_0=0.7528$,
$\bar{k}=5.0$. The total energy is $841 MeV$ and the RMS radius $0.437 fm$.   }
\end{figure}
\begin{figure}[h]
\caption{\label{stability1}
Time evolution of the quark distribution of a nucleon at rest in phase--space.
The
initialization is
described in the text. (time in $fm/c$).}
\end{figure}
\begin{figure}[h]
\caption{\label{stability2}
Time evolution of the scalar density (dashed) and $\sigma$--field (solid) for a
nucleon at
rest.
(time in $fm/c$) }
\end{figure}
\begin{figure}[h]
\caption{\label{coll1}
Phase--space evolution of the quark distribution in a collision of two nucleons
for a
bombarding
energy of $15 GeV$ in the lab.\ system. The time is given in $fm/c$. }
\end{figure}

\begin{figure}[h]
\caption{\label{coll2}
Time evolution of the scalar density and the $\sigma$--field for a collision of
two nucleons
with a
bombarding energy of $15 GeV$ in the lab.\ system. The time is given in
$fm/c$.}
\end{figure}
\begin{figure}[h]
\caption{\label{Fige1}
Time evolution of the energy for a collision with bombarding energy of $1 GeV$
in the
lab.\ system.
Total energy (solid), testparticle energy (dashed), $\sigma$--field kinetic
energy (dotted)
and $\sigma$--field potential energy (dashed--dotted).}
\end{figure}
\begin{figure}[h]
\caption{\label{Fige2}
Time evolution of the energy for a collisioon with bombarding energy of $15
GeV$ in the
lab.\
system.
Total energy (solid), testparticle energy (dashed), $\sigma$--field kinetic
energy (dotted)
and $\sigma$--field potential energy (dashed--dotted).}
\end{figure}

\end{document}